\documentclass[epj,final]{svjour}

\usepackage{latexsym}
\usepackage{graphics}
\usepackage{amsmath}

\newcommand {\bnabla} {{\vec{\nabla}}}

\newcommand {\bfp} {{\bf p}}
\newcommand {\bfq} {{\bf q}}
\newcommand {\bfr} {{\bf r}}
\newcommand {\bfv} {{\bf v}}

\newcommand {\bfE} {{\bf E}}
\newcommand {\bfY} {{\bf Y}}

\newcommand {\E} {\varepsilon}

\newcommand {\om} {\omega}
\newcommand {\Om} {\Omega}

\newcommand {\ee} {{\rm e}}

\newcommand {\pol} {{\rm pol}}

\renewcommand {\d} {{\rm d}}
\renewcommand {\i} {{\rm i}}
\renewcommand {\Im} {{\rm Im}}

\newcommand {\cF} {{\cal F}}

\newcommand {\limxi} {{\lim_{\xi \to 1}\,}}

\begin{document}

\title{Formalism of collective electron excitations in fullerenes}

\author{A. Verkhovtsev\inst{1,2}
\thanks{verkhovtsev@fias.uni-frankfurt.de}
\and A.V. Korol\inst{1} \and A.V. Solov'yov\inst{1}
\thanks{On leave from A.F. Ioffe Physical Technical Institute, St. Petersburg, Russia}
}                     
%
%
\institute{Frankfurt Institute for Advanced Studies, Goethe-Universit\"at, Ruth-Moufang-Str. 1, 60438 Frankfurt am Main, Germany
\and
St. Petersburg State Polytechnic University, Politekhnicheskaya ul. 29, 195251 St. Petersburg, Russia
}
%
\date{\today}
%
\abstract{
We present a detailed formalism for the description of collective electron excitations in fullerenes in the process of the electron inelastic scattering.
Considering the system as a spherical shell of a finite width, we show that the differential cross section is defined by three plasmon excitations, namely two coupled modes of the surface plasmon and the volume plasmon.
The interplay of the three plasmons appears due to the electron diffraction of the fullerene shell. Plasmon modes of different angular momenta provide dominating contributions to the differential cross section depending on the transferred momentum.
}
%
%
\maketitle
%
\section{Introduction}
\label{Intro}
In the present paper, we give a detailed theoretical explanation of the formation of various plasmon excitations in fullerenes caused by collision with fast electrons.
We demonstrate that a non-uniform electric field of a charged projectile causes variation of the electron density on the inner and the outer surfaces of the molecule as well as the volume density variation inside the fullerene shell.
The variation of electron density leads to the formation of three plasmon excitations, namely two surface plasmons and the volume plasmon.

Plasmon excitations represent the collective oscillation of electrons of an atomic system against the positively charged ions. This collective electronic motion appears when the system is exposed to an external field (the electromagnetic field or the electric field of a charged projectile). Collective excitations manifest themselves in the formation of giant resonances in the excitation spectrum of atomic systems. The existence of plasmon resonances is a general phenomenon occurring in various atomic systems, while a position of the resonance depends strongly on the type of the system. For instance, collective excitations in many-electron atoms have typical resonance frequencies of about 100 eV \cite{Connerade_GR}. Plasmon excitations in various atomic clusters have much lower frequencies, namely of the order of several eV for metal clusters \cite{Brechignac_1989_ChemPhysLett.164.433,Brack_1993_RevModPhys.65.677} and of several tens eV for fullerenes \cite{Hertel_1992_PhysRevLett.68.784,Keller_Coplan_1992_ChemPhysLett.193.89}.

Delocalized electrons of atomic clusters may form two different types of collective excitations, namely the surface and the volume plasmons \cite{deHeer_1993_RevModPhys.65.611,Kreibig_Vollmer}. The dipole surface plasmon is responsible for the formation of the giant resonance in photoabsorption
spectra of metal clusters \cite{Haberland,Ekardt} and fullerenes \cite{Hertel_1992_PhysRevLett.68.784}, and play also an important role in the process of inelastic scattering of electrons \cite{Solovyov_NATO_2001,Solovyov_review_2005_IntJModPhys.19.4143}.
The excitation of metal clusters revealed the existence of the volume plasmon which has a higher resonance frequency and is essential
for the formation of the electron impact ionization cross section \cite{Gerchikov_2000_PhysRevA.62.043201}.

Existence of the giant resonance in the excitation spectra of fullerenes at about 20 eV was predicted theoretically \cite{Bertsch_1991_PhysRevLett.67.2690} and then observed experimentally for the gas phase C$_{60}$ compounds studying the processes of photoionization \cite{Hertel_1992_PhysRevLett.68.784} and inelastic scattering of electrons \cite{Keller_Coplan_1992_ChemPhysLett.193.89}. Formation of the dipole plasmon resonance in the photoionization cross section of C$_{60}$ was studied theoretically within various models and approaches \cite{Ivanov_2001_JPhysB.34.L669,Madjet_2008_JPhysB.41.105101}.
Recent experiments on photoionization of neutral \cite{Reinkoester_2004_JPhysB.37.2135} and charged \cite{Scully_2005_PhysRevLett.94.065503} C$_{60}$ molecules revealed the existence of the second collective resonance at about 40 eV which later was associated \cite{Korol_AS_2007_PhysRevLett_Comment} with the second surface plasmon.

Theoretical investigations of the scattering of fast electrons on fullerenes \cite{Gerchikov_1997_JPhysB.30.4133,Gerchikov_1998_JPhysB.31.3065} revealed the existence of the diffraction phenomena arising in the scattering processes. The first experimental observation of the electron diffraction on C$_{60}$ was reported in Ref. \cite{Mikoushkin_1998_PhysRevLett.81.2707}. It was shown that plasmon modes of different angular momenta provide dominating contributions to the differential cross section at different electron scattering angles.
In the cited papers, the fullerene was modeled as an infinitely thin sphere and collective electron excitations were represented by the single surface plasmon.

In Ref. \cite{Connerade_AS_PhysRevA.66.013207} a hydrodynamic model was applied to describe the plasmon excitations formation mechanism. It was shown that no volume plasmons can be excited in the dipole-photon limit. Therefore, the volume plasmon can manifest itself only when the system interacts with a non-uniform external field, e.g. in collisions with charged particles.

The paper is organized as follows. Section \ref{Theory} is dedicated to the detailed description of the theoretical framework. We introduce the model of a fullerene in Sec. \ref{PlasmonExcitation}.
In Sec. \ref{Appendix_General_case}, we introduce general expressions based on classical electrodynamics and hydrodynamics describing collective electron excitations in a many-electron system.
In Sec. \ref{Spherical_system}, we derive a general expression for the electron density variation in an arbitrary spherically symmetric system exposed to an external field. In Sec. \ref{Appendix_Finite_width}, we apply this formalism to a fullerene considered within the introduced model.
In Sec. \ref{InelasticCS}, we derive general expressions for the inelastic scattering cross section.
In Sec. \ref{PRA}, we present the expressions for the volume and surface plasmon excitations obtained within the plasmon resonance approximation.
We discuss also the diffraction phenomena which manifest themselves in the contribution of terms of different multipolarity.
In Sec. \ref{LimitingCases}, we consider two limiting cases of the general equations presented in Sec. \ref{Theory}. We consider the ''metal cluster'' limit when the system is represented by a charged full sphere, thus the volume and the single surface plasmons appear in the system. We consider also the ''infinitely thin fullerene'' limit when collective excitations are described only by the single surface plasmon. The obtained expressions coincide with the results presented earlier in Refs. \cite{Gerchikov_2000_PhysRevA.62.043201,Gerchikov_1997_JPhysB.30.4133,Gerchikov_1998_JPhysB.31.3065}.
In Sec. \ref{Conclusion}, we draw the conclusions from this work.
In Appendix, 
we describe the transformation of general expressions obtained in Sec. \ref{Theory} in the case of a uniform external field. We show that no volume plasmon can appear in the system exposed to a uniform field and 
the photoionization cross section is defined then by the two surface plasmons.

The atomic system of units, $m_e = e = \hbar = 1$, is used throughout the paper.

\section{Theoretical framework}
\label{Theory}
\subsection{Model of a fullerene}
\label{PlasmonExcitation}

In this paper, we consider a fullerene $C_N$ as a spherically symmetric system where the charge is distributed homogeneously between two
concentric spheres \cite{Lambin_Lukas_1992_PhysRevB.46.1794,Oestling_1993_EurophysLett.21.539,Lo_Korol_AS_2007_JPhysB.40.3973}.
The width of the fullerene is $\Delta R = R_2 - R_1$ where $R_1$, $R_2$ are the inner and the outer radii of the molecule,
respectively.
The equilibrium electron density distribution $\rho_{0}(r)$ is expressed via the number $N$ of delocalized electrons
(four 2s$^2$2p$^2$ valence electrons from each carbon atom) and the fullerene volume $V$:

\begin{equation}
\rho_{0} = \left\{
\begin{array}{l l}
N/V         & \quad \textrm{for $R_1 \leq r \leq R_2$}
\\
0           & \quad \textrm{if otherwise} \ .
\end{array} \right.
\label{Notation.1}
\end{equation}

The volume of the fullerene shell can be expressed as
\begin{equation}
V = \frac{4\pi}{3} \Bigl( R_{2}^3 - R_{1}^3 \Bigr) = \frac{4\pi}{3} R_2^3 \Bigl( 1 - \xi^3 \Bigr) \ ,
\label{Notation.2}
\end{equation}
where $\xi = R_1/R_2 \leq 1$ is the ratio of the inner to the outer radii.

This model is applicable for any spherically symmetric system with an arbitrary value of the ratio $\xi$. Supposing $\xi = 0$, one obtains a model of a metal cluster, while the case $\xi = 1$ represents a fullerene modeled as an infinitely thin sphere.

When the fullerene is exposed to an external field, the electron density is modified, $\rho = \rho_0 + \delta \rho$.
Below, we present a detailed formalism of the electron density variation, $\delta \rho$.

\subsection{Variation of the electron density}
\subsubsection{General equations
\label{Appendix_General_case}}

Let us describe briefly a simplified version of the approach introduced in Ref. \cite{Connerade_AS_PhysRevA.66.013207}.
The presented formalism describes the behavior of collective electronic excitations and is based on classical electrodynamics and hydrodynamics.

Let $\rho_0(\bfr)$ denote the stationary distribution of the electron charge in the point $\bfr$.
The variation of the electron density, $\delta \rho(\bfr,t)$, depends on the position $\bfr$ and time $t$.
Therefore, the total electron density is introduced as:
\begin{equation}
\rho(\bfr,t)=\rho_0(\bfr) + \delta \rho(\bfr,t) \ .
\label{BasicEquations.1}
\end{equation}

Following \cite{Connerade_AS_PhysRevA.66.013207}, let us describe the collective motion of the electron density using the Euler equation and the equation of continuity.

The Euler equation couples the acceleration, $\d \bfv(\bfr,t)/\d t$, of the electron density with the total electric field $\bfE$ acting on the system at the point $(\bfr,t)$:
\begin{equation}
{\d \bfv(\bfr,t) \over \d t} = \bfE(\bfr,t) \ .
\label{BasicEquations.2}
\end{equation}
The electric field $\bfE$ includes both the external field acting on the
system and the polarization contribution due to the variation of electron density $\delta \rho(\bfr,t)$:
\begin{equation}
{\d \bfv(\bfr,t) \over \d t} = - \bnabla \phi(\bfr,t)
- \bnabla
\int
{\d \bfr^{\prime}\delta\rho(\bfr^{\prime},t)\over |\bfr-\bfr^{\prime}|} \ ,
\label{BasicEquations.3}
\end{equation}
where $\phi(\bfr,t)$ is the scalar potential of the external field.

Introducing (\ref{BasicEquations.3}) in (\ref{BasicEquations.2})
and evaluating the full time derivative of the vector $\bfv$
one obtains
\begin{eqnarray}
{\partial \bfv(\bfr,t) \over \partial t}
&+&
\Bigl(\bfv(\bfr,t) \cdot \bnabla\Bigr)\,\bfv(\bfr,t)
= \nonumber \\
&-&  \Bigl(\bnabla \phi(\bfr,t)\Bigr)
- \bnabla
\int
\d \bfr^{\prime} {\delta\rho(\bfr^{\prime},t)\over |\bfr-\bfr^{\prime}|} \ .
\label{BasicEquations.4}
\end{eqnarray}

The potential of the external field is assumed to satisfy the wave equation
and has the monochromatic dependence on $t$:
\begin{equation}
 \phi(\bfr,t) = \ee^{\i \om t}\, \phi(\bfr) \ ,
\label{BasicEquations.5a}
\end{equation}
where $\phi(\bfr)$ satisfies the equation
\begin{equation}
\Delta \phi(\bfr) + q^2\, \phi(\bfr) =0
\label{BasicEquations.5b}
\end{equation}
with $q$ is the wave vector.

The motion of electron density in the system obeys the equation of continuity, which reads
\begin{equation}
{\partial \rho(\bfr,t) \over \partial t}
+ \bnabla \cdot \Bigl(\rho(\bfr,t)\, \bfv(\bfr,t)\Bigr) = 0 \ .
\label{BasicEquations.5}
\end{equation}

Eqs. (\ref{BasicEquations.4}) and (\ref{BasicEquations.5}), being solved simultaneously, determine the variation of electron density $\delta \rho(\bfr,t)$ as well as its velocity $\bfv(\bfr,t)$.

Let us estimate the relative value of the first and the second terms on the left-hand side of Eq. (\ref{BasicEquations.4}).
To make an estimation, let us consider the one-dimensional case, and let $E\propto \exp(\i\om t)$. Thus, Eq. (\ref{BasicEquations.4}) can be written in the simplified form:
\begin{equation}
\dot{v} + v\,{\partial v \over \partial x} =  E_0 \ee^{\i\om t} \ .
\label{Perturbation.my.1}
\end{equation}
Here ${\partial v / \partial x}\sim v/R$ where $R$ is the size of the
system.
Supposing that the term $\dot{v}$ dominates,
one finds that $v\sim (1/\om)E_0\exp(\i\om t)$.
Hence, from the condition $\dot{v}\gg v^2/R$ one can obtain the criterion $E\ll \om^2 R$.
Below, we assume that this condition is fulfilled and neglect the second term on the left-hand side of Eq. (\ref{BasicEquations.4}), which means physically that the external field causes only a small spatial inhomogeneity in the electron density distribution within the system.

Let us seek the solutions of Eqs. (\ref{BasicEquations.4}) and (\ref{BasicEquations.5}) in the following form:
\begin{eqnarray}
\left\{
\begin{array}{l l}
\delta\rho(\bfr,t) = \delta\rho(\bfr)\, \ee^{\i \om t} \vspace{0.2cm} \\
\bfv(\bfr,t) = \bfv(\bfr)\, \ee^{\i \om t} \ .
\end{array}
\right.
\label{Perturbation.1}
\end{eqnarray}

Substituting these expressions into Eqs. (\ref{BasicEquations.4}) and (\ref{BasicEquations.5}) and performing some transformations with the simultaneous use of Eq. (\ref{BasicEquations.5b}) and
$\Delta |\bfr - \bfr'|^{-1} = -4\pi \delta(\bfr - \bfr')$, one derives a set of the following linear equations:
\begin{equation}
\bfv(\bfr) = \frac{\i}{\om}
\left[ \bnabla \phi(\bfr) +
\bnabla\int
{\delta\rho(\bfr^{\prime}) \over |\bfr-\bfr^{\prime}|}\,\d \bfr^{\prime}
\right]
\label{Perturbation.2}
\end{equation}
and
\begin{align}
& \Bigl(\om^2 - 4\pi\rho_0(\bfr)\Bigr) \delta\rho(\bfr) +
\bnabla\rho_0(\bfr) \cdot
\bnabla\int
{\delta\rho(\bfr^{\prime}) \over |\bfr-\bfr^{\prime}|}\,\d \bfr^{\prime}
\nonumber\\
& \qquad
=
q^2\,\rho_0(\bfr)\,\phi(\bfr)
-\bnabla\rho_0(\bfr) \cdot \bnabla \phi(\bfr) \ .
\label{Perturbation.12a}
\end{align}

\subsubsection{Arbitrary spherically symmetric system
\label{Spherical_system}}

Eqs. (\ref{Perturbation.2}) and (\ref{Perturbation.12a}) describe the dynamics of electron density under the action of an external field.
In the case of the spherically symmetric density distribution, $\rho_0(\bfr) = \rho_0(r)$, one can exclude angular variables from Eq. (\ref{Perturbation.2}) and (\ref{Perturbation.12a}).

Let us expand functions $\phi(\bfr)$, $\delta\rho(\bfr)$ and $|\bfr - \bfr'|^{-1}$ into spherical harmonics:
\begin{eqnarray}
\left\{
\begin{array}{l l}
\displaystyle{ \phi(\bfr) = \sum\limits_{lm} \phi_l(r) Y_{lm}(\bfr) }
\\
\displaystyle{ \delta\rho(\bfr) = \sum\limits_{lm} \delta\rho_l(r) Y_{lm}(\bfr) } \\
\displaystyle{
{1 \over |\bfr - \bfr^{\prime}|} = \sum_{lm} \frac{4\pi}{\Pi_l^2}  b_l(r,r^{\prime})\,
Y_{lm}(\bfr)Y_{lm}^{*}(\bfr^{\prime}) } \ ,
\end{array}
\right.
\label{CS2002.Spherically.3}
\end{eqnarray}
where the function $b_l(r,r^{\prime})$ is defined as follows:
\begin{align}
& b_l(r,r^{\prime})={r_{<}^l \over r_{>}^{l+1}}
=
{r^l \over (r^{\prime})^{l+1}}\, \Theta(r^{\prime}-r)
+
{(r^{\prime})^l \over r^{l+1}}\, \Theta(r-r^{\prime}) \ ,
\label{CS2002.Spherically.3a}
\end{align}
$\Theta(x)$ is the Heaviside step function defined as
\begin{equation}
\Theta(x) = \left\{
\begin{array}{l l}
0 \ , \quad         & x < 0
\\
1 \ , \quad         & x \ge 0
\end{array} \right. ,
\label{Notation.101}
\end{equation}
and the notation $\Pi_l = \sqrt{2l+1}$ is introduced.

Using the last two expressions in Eq. (\ref{CS2002.Spherically.3}), one gets
\begin{align}
\int
{\delta\rho(\bfr^{\prime}) \over |\bfr-\bfr^{\prime}|}
\d \bfr^{\prime}
=
\sum\limits_{lm}
{4\pi \over \Pi_l^2} Y_{lm}(\bfr)
\int\limits_0^{\infty}
r^{\prime 2} b_l(r,r^{\prime})
\delta\rho_l(r^{\prime}) \d r^{\prime} \ .
\label{CS2002.Spherically.4}
\end{align}

To expand the terms containing the operator $\bnabla$, the following general expression is used
(see Ref. \cite{Varshalovich}):
\begin{eqnarray}
\bnabla \Bigl(f(r)\, Y_{lm}(\bfr)\Bigr)
&=&
{\d f(r)\over \d r}\,
{\bfY}_{lm}^{(-1)}(\bfr) \nonumber \\
&+&
\sqrt{l(l+1)}\,
{f(r)\over r}\,
\bfY_{lm}^{(1)}(\bfr)
\label{CS2002.Spherically.5} \ ,
\end{eqnarray}
where ${\bfY}_{lm}^{(-1)}(\bfr)$ and $\bfY_{lm}^{(1)}(\bfr)$ are the longitudinal and the transverse vector spherical harmonics, respectively. The definition of these functions can be found in Ref. \cite{Varshalovich}.

Hence
\begin{align}
&\displaystyle{
\bnabla \int
{\delta\rho(\bfr^{\prime}) \over |\bfr-\bfr^{\prime}|}\,
\d \bfr^{\prime}
=
\sum\limits_{lm}
{4\pi \over \Pi_l^2}
\int\limits_0^{\infty}\d r^{\prime}\,r^{\prime\, 2}\,
\delta\rho_l(r^{\prime}) \, }
\nonumber \\
&\times
\displaystyle{
\left[ { \partial b_l(r,r^{\prime})\over \partial r}\,{\bfY}_{lm}^{(-1)}(\bfr)
+
\sqrt{l(l+1)}\,
{b_l(r,r^{\prime})\over r}\,
\bfY_{lm}^{(1)}(\bfr)
\right] }
\label{CS2002.Spherically.6a}
\end{align}
and
\begin{eqnarray}
\bnabla \phi(\bfr)
&=&
 \sum\limits_{lm}
\left[ { \d \phi_l(r)\over \d r}\,{\bfY}_{lm}^{(-1)}(\bfr) \right.
\nonumber \\
&+&
\left. \sqrt{l(l+1)}\,
{\phi_l(r)\over r}\,
\bfY_{lm}^{(1)}(\bfr)
\right] \ .
\label{CS2002.Spherically.6}
\end{eqnarray}

Accounting for the identities (see Ref. \cite{Varshalovich})
\begin{align}
&\bfY_{lm}^{(-1)}(\bfr)\cdot\bfY_{l'm'}^{(1)}(\bfr)=0  \\
&\bfY_{lm}^{(-1)}(\bfr)\cdot\bfY_{l'm'}^{(-1)}(\bfr)=Y_{lm}(\bfr)Y_{l'm'}(\bfr) \ ,
\end{align}
Eq. (\ref{Perturbation.12a}) is transformed to the following expression:
\begin{align}
& \sum_{l m}
Y_{lm}(\bfr)
\left[
\Biggl( \om^2 - 4\pi \rho_0(r)\Biggr) \delta\rho_l(r)
\right. \nonumber \\
& \qquad \qquad + \left.
4\pi {\rho_0^{\prime}(r) \over \Pi_l^2}
\int\limits_0^{\infty} g_l(r,r^{\prime})\,\delta\rho_l(r^{\prime}) \d r^{\prime}
\right]
\nonumber\\
\quad
&=
 \sum_{l m}
Y_{lm}(\bfr)
\left[
q^2\,\rho_0(r)\,\phi_l(r)
-
\rho_0^{\prime}(r)\, \phi_l^{\prime}(r)
\right] \ ,
\label{CS2002.Spherically.9a}
\end{align}
where the function $g_l(r,r^{\prime})$ is related to $b_l(r,r^{\prime})$ defined in (\ref{CS2002.Spherically.3a}):
\begin{align}
&g_l(r,r^{\prime}) = r^{\prime\, 2}\, { \partial b_l(r,r^{\prime})
\over \partial r}  \\
&=
l\,\left({r \over r^{\prime}}\right)^{l-1}\, \Theta(r^{\prime}-r)
-
(l+1)
\left({r^{\prime} \over r}\right)^{l+2}\, \Theta(r-r^{\prime}) \nonumber \ .
\label{CS2002.Spherically.9b}
\end{align}

Multiplying both sides of Eq. (\ref{CS2002.Spherically.9a})
by $Y_{lm}^{*}(\bfr)$ and carrying out the integration over spherical angles
of the vector $\bfr$, one obtains a general equation for the variation of electron density in an arbitrary spherically symmetric system:
\begin{eqnarray}
\Bigl( \om^2 &-& 4\pi\rho_0(r) \Bigr) \delta\rho_l(r)
+
4\pi \frac{\rho_0'(r)}{\Pi_l^2}
\int\limits_0^{\infty} g_l(r,r') \delta\rho_l(r') \d r'
\nonumber \\
&=&
q^2 \rho_0(r) \phi_l(r) - \rho_0'(r) \phi_l'(r) \ ,
\label{CS2002.Spherically.9c}
\end{eqnarray}
where $\phi_l(r)$ is the scalar potential and $\om$ is the frequency of the external field.

\subsubsection{Fullerene as spherical shell of a finite width
\label{Appendix_Finite_width}}

In the case of the fullerene model described in Sec. \ref{PlasmonExcitation},
the equilibrium electron density distribution, $\rho_0(r)$, is constant within the interval $R_1 < r < R_2$
and equals to zero if otherwise:
\begin{equation}
\rho_0(r) = \rho_0\, \Theta(r - R_1)\Theta(R_2 - r) \ ,
\label{Appendix.01}
\end{equation}
where $\rho_0$ is defined by Eq. (\ref{Notation.1}). The derivative of the function $\rho_0(r)$ is given by
\begin{equation}
\rho_0'(r) = \rho_0\, \Bigl( \delta(r - R_1) - \delta(r - R_2) \Bigr) \ ,
\label{Appendix.02}
\end{equation}
where $\delta(x)$ is the delta-function.

\begin{figure}[h]
\centering
\resizebox{0.97\columnwidth}{!}{
\includegraphics{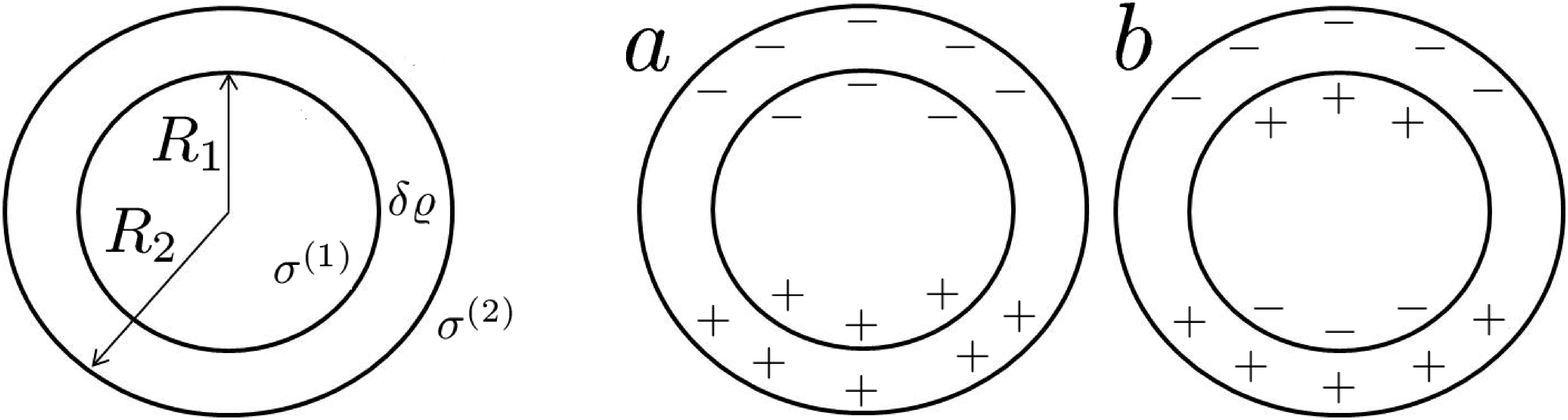}
}
\caption{Left panel: Representation of a fullerene as a spherical shell of a width $R_2 - R_1$. Variation of the surface
charge densities, $\sigma^{(1,2)}$, and the volume charge density, $\delta\varrho$, is also shown. Right panel: Representation
of the symmetric (a) and the antisymmetric (b) modes of the surface plasmon.}
\label{figure1}
\end{figure}

Then, the solution of Eq. (\ref{CS2002.Spherically.9c}) for the fullerene is sought in the following form:
\begin{eqnarray}
\delta\rho_l(r) &=& \delta\varrho_l(r)\Theta(r - R_1)\Theta(R_2 - r)  \nonumber \\
&+& \sigma_l^{(1)} \delta(r - R_1) + \sigma_l^{(2)} \delta(r - R_2) \ ,
\label{Appendix.05}
\end{eqnarray}
where $\delta\varrho_l(r)$ describes the volume density variation arising inside the fullerene shell, and $\sigma_l^{(1,2)}$ are variations of the surface charge densities at the inner and the outer surfaces of the shell, respectively (see the left panel of Fig. \ref{figure1}). The volume density variation causes the formation of the volume plasmon, while the variations of the surface densities correspond to two surface plasmon modes.

Using (\ref{Appendix.01}),(\ref{Appendix.02}) and (\ref{Appendix.05}) in Eq. (\ref{CS2002.Spherically.9c}) and carrying out algebraic transformations, one derives:
\begin{align}
& \bigl( w - 1 \bigr) \delta\varrho_l(r) \Theta(r - R_1)\Theta(R_2 - r)  \nonumber \\
& \quad + \biggl(
w \sigma_l^{(1)} +
 I_l^{(1)} -
        \frac{l+1}{\Pi_l^2} \sigma_l^{(1)} + \frac{l}{\Pi_l^2} \sigma_l^{(2)}\, \xi^{l-1} \biggr)
 \delta(r - R_1)
\nonumber \\
& \quad + \biggl(
w \sigma_l^{(2)} +
 I_l^{(2)} +
        \frac{l+1}{\Pi_l^2} \sigma_l^{(1)} \xi^{l+2} - \frac{l}{\Pi_l^2}\sigma_l^{(2)} \biggr)
 \delta(r - R_2)
\nonumber \\
& = q^2 \frac{\phi_l(r)}{4\pi} \Theta(r - R_1)\Theta(R_2 - r)
\nonumber \\
& \quad - \frac{1}{4\pi} \Bigl( \phi_l'(R_1)\delta(r - R_1) - \phi_l'(R_2)\delta(r - R_2) \Bigr) \ ,
\label{Appendix.06}
\end{align}
where the following notations are introduced:
\begin{equation}
w = \frac{\om^2}{\om_p^2} \ ,
\label{Appendix.09}
\end{equation}
\begin{eqnarray}
\left\{
\begin{array}{l l}
\displaystyle{ I_l^{(1)} = \frac{l}{\Pi_l^2} R_1^{l-1} \int\limits_{R_1}^{R_2} \frac{\delta\varrho_l(x)}{x^{l-1}}\, \d x }
\\
\displaystyle{ I_l^{(2)} = \frac{l+1}{\Pi_l^2} \frac{1}{R_2^{l+2}}
\int\limits_{R_1}^{R_2} x^{l+2} \delta\varrho_l(x) \d x } \ .
\end{array} \right.
\label{Appendix.12}
\end{eqnarray}
Parameter $\om_{p}$ is the volume plasmon frequency associated with the density $\rho_{0}$. Neglecting the dispersion, the volume plasmon frequency has a constant value and is defined as:
\begin{equation}
\om_{p}^2 = 4\pi \rho_{0} = \frac{3N}{R_2^3 (1-\xi^3)} \ .
\label{Notation.4}
\end{equation}

Matching the terms of different types on the right- and the left-hand side of Eq. (\ref{Appendix.06}), one obtains three equations:
one for the volume plasmon and the other two - for the surface plasmons.

The solution corresponding to the volume density variation reads as:
\begin{equation}
\delta\varrho_l(r) = \frac{q^2}{w - 1} \frac{\phi_l(r)}{4\pi}  \ .
\label{Appendix.08}
\end{equation}

The total density variation due to the surface plasmons is
\begin{equation}
\sigma_l(r) = \sigma_l^{(1)} \delta(r - R_1) + \sigma_l^{(2)} \delta(r - R_2) \ ,
\label{Appendix.10}
\end{equation}
where the quantities $\sigma_l^{(1)}$ and $\sigma_l^{(2)}$ satisfy the following system of coupled equations:
\begin{eqnarray}
\left\{
\begin{array}{l l}
\displaystyle{ \biggl( w - \frac{l+1}{\Pi_l^2} \biggr) \sigma_l^{(1)} +
\frac{l}{\Pi_l^2}\xi^{l-1}\sigma_l^{(2)} = F_1 }
\vspace{0.2cm} \\
\displaystyle{ \frac{l}{\Pi_l^2}\xi^{l+2}\sigma_l^{(1)} +
\biggl( w - \frac{l}{\Pi_l^2} \biggr) \sigma_l^{(2)} = F_2 }
\end{array} \right. \ ,
\label{Appendix.11}
\end{eqnarray}
where
\begin{equation}
F_{1,2} = \mp \frac{\phi_l'(R_{1,2})}{4\pi} - I_l^{(1,2)} \ .
\label{Appendix.15}
\end{equation}

Using the expression (\ref{Appendix.08}) for the volume density variation, the functions $I_l^{(1)}$ and $I_l^{(2)}$ can be rewritten in the following form:
\begin{eqnarray}
\left\{
\begin{array}{l l}
\displaystyle{ I_l^{(1)} = \frac{q^2}{w - 1} \frac{l}{\Pi_l^2} \int\limits_{R_1}^{R_2} \frac{R_1^{l-1}}{x^{l-1}} \frac{\phi_l(r)}{4\pi}
\d x }
\\
\displaystyle{ I_l^{(2)} = \frac{q^2}{w - 1} \frac{l+1}{\Pi_l^2} \int\limits_{R_1}^{R_2} \frac{x^{l+2}}{R_2^{l+2}} \frac{\phi_l(r)}{4\pi} \d x } \ .
\end{array} \right.
\label{Appendix.12}
\end{eqnarray}
The determinant of the system (\ref{Appendix.11}) is
\begin{equation}
\Delta = (w - w_{1l})(w - w_{2l}) \ ,
\label{Appendix.13}
\end{equation}
where $w_{1l}$ and $w_{2l}$
are the roots of the secular equation $\Delta = 0$:
\begin{eqnarray}
\left\{
\begin{array}{l l}
\displaystyle{ w_{1l} = \frac12 \Bigl(1 -  \frac{1}{2l+1} \sqrt{1 + 4l(l+1)\xi^{2l+1}}\, \Bigr) }
\vspace{0.2cm} \\
\displaystyle{ w_{2l} = \frac12 \Bigl(1 +  \frac{1}{2l+1} \sqrt{1 + 4l(l+1)\xi^{2l+1}}\, \Bigr) }
\end{array} \right. \ .
\label{MultipoleVariation.2a}
\end{eqnarray}

Solutions of the system (\ref{Appendix.11}) are given by the following expression:
\begin{eqnarray}
\left\{
\begin{array}{l l}
\displaystyle{ \sigma_l^{(1)} =
\frac{1}{\Delta} \left[ F_1w - \frac{l}{\Pi_l^2} \bigl( F_1 + \xi^{l-1}F_2 \bigr) \right]
\equiv \frac{\cF_1}{\Delta} }
\vspace{0.3cm} \\
\displaystyle{ \sigma_l^{(2)} =
\frac{1}{\Delta} \left[ F_2w - \frac{l+1}{\Pi_l^2} \bigl( F_2 + \xi^{l+2}F_1 \bigr) \right]
\equiv \frac{\cF_2}{\Delta} } \ .
\end{array} \right.
\label{Appendix.14}
\end{eqnarray}

Variation of the surface charge densities, $\sigma_l^{(1,2)}$, results in the formation of two coupled modes of surface plasmon oscillations \cite{Lambin_Lukas_1992_PhysRevB.46.1794,Oestling_1993_EurophysLett.21.539,Lo_Korol_AS_2007_JPhysB.40.3973}.
Frequencies of the symmetric, $\om_{1l}$, and the antisymmetric, $\om_{2l}$, surface plasmons of multipolarity $l$ are given
by the expression \cite{Oestling_1993_EurophysLett.21.539,Lo_Korol_AS_2007_JPhysB.40.3973}:
\begin{eqnarray}
\om_{jl}^2 = w_{jl} \om_p^2 \quad (j = 1,2) \ .
\label{Appendix.17}
\end{eqnarray}
In the symmetric mode the charge densities on
the two surfaces oscillate in phase, while in the antisymmetric mode they are out of phase (see the right panel of Fig.~\ref{figure1}).

Using Eqs. (\ref{Appendix.08}) and (\ref{Appendix.14}) in (\ref{Appendix.05}), one obtains the expression which defines the multipole variation of electron density in a spherically symmetric fullerene of a finite width under the action of the multipole component $\phi_l(r)$ of the external field:
\begin{eqnarray}
 \delta\rho_l(r) &=& \displaystyle{ \frac{q^2}{w - 1} \frac{\phi_l(r)}{4\pi} } \Theta(r - R_1)\Theta(R_2 - r)  \nonumber \\
&+& \frac{\cF_1}{\Delta} \delta(r - R_1) + \frac{\cF_2}{\Delta} \delta(r - R_2)  \ .
\label{Appendix.16}
\end{eqnarray}

\subsection{Inelastic scattering of an electron\label{InelasticCS}}

\begin{figure}[h]
\centering
\resizebox{0.6\columnwidth}{!}{
\includegraphics{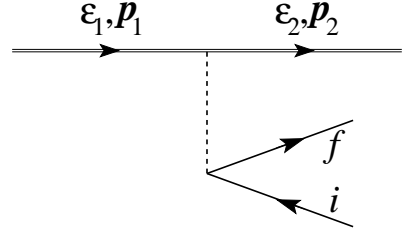}
}
\caption{Diagrammatic representation of the inelastic scattering process. The projectile electron goes from the initial electron
state $(\E_1, \bfp_1)$ to the final state $(\E_2, \bfp_2)$ while the fullerene is ionized from the initial state $i$ ($\E_i$)
to the final state $f$ ($\E_f$).}
\label{figure2}
\end{figure}

In the process of inelastic scattering the projectile electron undergoes the transition from the initial electron state
$(\E_1, \bfp_1)$ to the final state $(\E_2, \bfp_2)$ which is accompanied by the ionization (or, excitation) of a fullerene from the initial
state $i$ with the energy $\E_i$ to the final state $f$ with $\E_f$.
Diagrammatic representation of the process is given in Fig. \ref{figure2}.

The matrix element, $M$, which defines the amplitude of the inelastic scattering is given by
\begin{align}
M &= \left\langle f,2 \left| \sum_{a} \frac{1}{|{\bf r} - {\bf r}_a|} \right| 1,i \right\rangle
\nonumber \\
&= \sum_a \int \psi_2^{(-)*}({\bf r}) \psi_f^*(\{{\bf r}_a\}) \frac{1}{|{\bf r} - {\bf r}_a|} \nonumber \\
&\qquad\quad \times \psi_i(\{{\bf r}_a\}) \psi_1^{(+)}({\bf r}) \{ \d {\bf r}_a \} \d {\bf r} \ ,
\end{align}
where
$\{ {\bf r}_a \} = {\bf r}_1 \dots {\bf r}_N$ are the position vectors of the delocalized electrons in the fullerene, ${\bf r}$ is the position vector of the
projectile, $\psi_1^{(+)}({\bf r})$ and $\psi_2^{(-)}({\bf r})$ stand for the initial- and the final state wave functions of the
projectile, respectively.
Superscripts $(+)$ and $(-)$ indicate that asymptotic behavior of the wave functions is 'plane wave + outgoing spherical wave' and 'plane wave + incoming wave', respectively.

The matrix element can be written as follows:
\begin{equation}
M =
\int {4\pi \over q^2} {\d \bfq \over (2\pi)^3}
\left\langle 2\left|\ee^{-\i \bfq\cdot\bfr} \right| 1\right\rangle
\left\langle f\left|\sum_{a} \ee^{\i \bfq\cdot\bfr_a} \right| i\right\rangle \ ,
\label{first_Born}
\end{equation}
where $\bfq = \bfp_1-\bfp_2$ is the transferred momentum.

In the present paper, we consider the collision process of C$_{60}$ molecules with fast electrons. Since the collision velocity is larger than the characteristic velocities of delocalized electrons in the fullerene, the first Born approximation is applicable \cite{Gerchikov_1997_JPhysB.30.4133}. Within this approximation the initial and the final states of the incident electron can be described by plane waves:
\begin{equation}
\psi_1^{(+)}({\bf r}) = {\rm e}^{\i {\bf p}_1\cdot{\bf r}} \ , \quad \quad
\psi_2^{(-)}({\bf r}) = {\rm e}^{\i {\bf p}_2\cdot{\bf r}} \ .
\end{equation}
Within the framework of the plane-wave first Born approximation
the amplitude of the process reduces to
\begin{equation}
M = \frac{4\pi}{q^2}
\left\langle f\left|\sum_{a} \ee^{\i \bfq\cdot\bfr_a} \right| i\right\rangle_{\bfq = {\bf p}_1 - {\bf p}_2} \ .
\label{first_Born_2}
\end{equation}
The magnitude of $q^2$ is related to $p_{1,2}$ and the scattering angle $\theta = \widehat{\bfp_1 \, \bfp_2}$ via:
\begin{equation}
q^2 = p_1^2 + p_2^2 -2p_1p_2 \cos\theta \approx p_1^2 \theta^2 \ .
\end{equation}
The final approximate equality is valid when $p_1 \approx p_2$ and the scattering angle is small, $\theta \ll 1$ rad.

Performing the multipole expansion of the exponential factors in (\ref{first_Born_2}) (see, e.g., \cite{Varshalovich}), one obtains:
\begin{equation}
M = 4\pi
\sum_{lm}
\i^l\, Y_{lm}^{*}(\bfq)
\left\langle f\left|
\sum_{a} \phi_{l}(r_a) Y_{lm}(\bfr_a)
\right| i\right\rangle \ ,
\label{InelasticCS.5}
\end{equation}
where we introduced the following notation:
\begin{equation}
\phi_{l}(r) = 4\pi {j_l(qr) \over q^2}
\label{Equation_01}
\end{equation}
and $j_l$ is the spherical Bessel function.

Let us consider a general expression for the cross section of the scattering process:
\begin{equation}
\d \sigma = \frac{2\pi}{p_1}\, \delta(\om_{fi} - \om)
\sum_{\pol_f} \overline{\sum_{\pol_i}}
\left|M\right|^2
{\d \bfp_2 \over (2\pi)^3}\, \d \rho_f \ ,
\label{InelasticCS.6}
\end{equation}
where $\om = \E_1 - \E_2$ is the energy transfer, $\om_{fi} = \E_f-\E_i$ and $\om = \om_{fi}$ due to the energy conservation
law.
The sign $\sum_{{\rm pol}_f}$ denotes the summation over the projection of the final state $f$ orbital momentum, whereas
$\overline{\sum}_{{\rm pol}_i}$ denotes the averaging over the projections of the initial state orbital momentum,
$\d \rho_f$ is the density of the fullerene final states.

Substituting the scattering amplitude (\ref{InelasticCS.5}) into Eq. (\ref{InelasticCS.6}), one derives the triply differential cross section:
\begin{eqnarray}
{\d^3 \sigma \over \d \E_2 \d \Om_{\bfp_2}} = \frac{1}{\pi}\frac{p_2}{p_1}
\sum_{lm} \int
\left| \left\langle f \left|
\sum_{a} V_{lm}(\bfr_a)\right| i\right\rangle\right|_{\rm pol}^2 \nonumber \\
\times \delta(\om_{fi} - \om) \, \d \rho_f \ ,
\label{InelasticCS.12}
\end{eqnarray}
where
\begin{equation}
V_{lm}(\bfr) = \phi_{l}(r) Y_{lm}(\bfr)
\label{Equation_02}
\end{equation}
is the multipolar potential of the projectile electron, $\d \Om_{\bfp_2}$ denotes the differentiation over the solid angle
of the scattered electron and sign $\int \d \rho_f$ means the summation over the final states (which includes the summation over the discrete spectrum and the integration over the continuous spectrum).

\subsection{Plasmon resonance approximation}\label{PRA}

Let us consider the inelastic scattering cross section within the plasmon resonance approximation \cite{Gerchikov_1997_JPhysB.30.4133,Gerchikov_1998_JPhysB.31.3065}. It relies on the assumption that collective plasmon excitations give the main contribution to the cross section in the vicinity of the giant resonance. Hence, one can neglect single-particle excitations when calculating the matrix element in Eqs. (\ref{InelasticCS.6}) and (\ref{InelasticCS.12}).

According to the Kubo linear response theory \cite{Gerchikov_1997_JPhysB.30.4133,Kubo_lin_response_1962}, the integral on the right-hand side of Eq. (\ref{InelasticCS.12}) can be related to the density variation $\delta \rho_l$ (\ref{Appendix.16}),
so one can perform the following substitution:
\begin{eqnarray}
\int \left|\left\langle f\left|
\sum_{a} V_{lm}(\bfr_a)\right| i\right\rangle\right|_{\rm pol}^2
\delta(\om_{fi} - \om) \d \rho_f \ \
\longrightarrow \nonumber \\
\frac{1}{\pi}\, \Im \int V_{lm}^*(\bfr) \delta\rho_l(\om,q;\bfr) \d \bfr \ .
\label{InelasticCS.13}
\end{eqnarray}
Here $\delta\rho_l(\om,q;\bfr)$ is a partial density variation due to the exposure of the system to the multipolar potential
$V_{lm}(\bfr)$.
In a general case, this variation depends on the frequency $\om$, transferred momentum $q$ and the position vector $\bfr$ as
well.

Using (\ref{InelasticCS.13}) in (\ref{InelasticCS.12}), the triply differential cross section acquires the form:
\begin{equation}
\frac{\d^3\sigma}{\d\E_2 \d\Om_{\bfp_2}} = \frac{1}{\pi^2} \frac{p_2}{p_1} \sum_{l} \Im \Bigl[ I_l(\om,q) \Bigr] \ ,
\label{InelasticCS.14}
\end{equation}
where
\begin{equation}
I_l(\om,q) = \Pi_l^2 \int V_{lm}^*(\bfr) \delta\rho_l(\om,q;\bfr) \d\bfr
\label{InelasticCS.15}
\end{equation}
and the notation $\Pi_l = \sqrt{2l+1}$ is used.

Using the multipole variation of the electron density
$\delta\rho_l(\om,q;\bfr)$ defined by Eq. (\ref{Appendix.16}) as well as the notations (\ref{Equation_01}) and (\ref{Equation_02}), one can write
\begin{equation}
I_l(\om,q) = I_l^{(v)}(\om,q) + I_l^{(s_1)}(\om,q) + I_l^{(s_2)}(\om,q) \ ,
\end{equation}
where
\begin{eqnarray}
\left\{
\begin{array}{l l}
\displaystyle{ I_l^{(v)}(\om,q) = \frac{4\pi}{q^2} \Pi_l^2 \int\limits_0^{\infty} r^2 j_l(qr)
\delta\varrho_l(\om,q;r) \d r } \\
\displaystyle{ I_l^{(s_j)}(\om,q) = \frac{4\pi}{q^2} \Pi_l^2 \int\limits_0^{\infty} r^2 j_l(qr)
\sigma_l^{(j)}(\om,q;r) \d r  } \ ,
\end{array} \right.
\end{eqnarray}
and $j = 1,2$. The term $\delta\varrho_l(\om,q;r)$ is the electron density variation associated with the volume plasmon
(see Eq. (\ref{Appendix.08})), $\sigma_l^{(1)}$ and $\sigma_l^{(2)}$ describe the density variation at the inner and the outer fullerene surfaces (see Eq. (\ref{Appendix.14})).

Performing some transformations, we come to the formula for the differential inelastic scattering cross section
with no damping of plasmon oscillations:
\begin{align}
& \frac{\d^3\sigma}{\d\E_2 \d\Om_{{\bfp}_2}} = \frac{2}{\pi} \frac{R_2}{q^4} \frac{p_2}{p_1} \nonumber \\
& \times  \Im \sum_{l}
\Biggl[ \frac{\om_p^2\,V_l(q)}{\om^2-\om_p^2} +
\frac{\om_{1l}^2\,S_{1l}(q)}{\om^2-\om_{1l}^2} + \frac{\om_{2l}^2\,S_{2l}(q)}{\om^2-\om_{2l}^2}  \Biggr] \ ,
\label{PRA.01}
\end{align}
where $\om_p$ is the volume plasmon frequency defined by Eq. (\ref{Notation.4}), $\om_{1l}$ and $\om_{2l}$ are the frequencies
of the symmetric and antisymmetric surface plasmons of multipolarity $l$ defined by Eq. (\ref{Appendix.17}).
Functions $V_l(q)$, $S_{1l}(q)$ and $S_{2l}(q)$ are defined as follows:

\begin{eqnarray}
V_l(q) &=& \Pi_l^2 \Biggl[
z_2^2 \biggl( j_{l}^2(z_2) + j_{l}^{\prime2}(z_2) - \xi^3\Bigl( j_{l}^2(z_1) + j_{l}^{\prime2}(z_1)\Bigr) \biggr)
\nonumber \\
&+& 3z_2 \Bigl( j_{l}(z_2)j_{l}^{\prime}(z_2) - \xi^2 j_{l}(z_1)j_{l}^{\prime}(z_1) \Bigr)
\nonumber \\
&-& l(l+1)\, \Bigl( j_{l}^2(z_2) -  \xi\,j_{l}^2(z_1)\Bigr)
\nonumber\\
&-& \frac{2 \Pi_l^2 }{ 1 - \xi^{2l+1} }
\Bigl( (1-a_2)\,j_{l}^2(z_2) + \xi(1-a_1)\, j_{l}^2(z_1) \Bigr)
\nonumber \\
&+& \frac{4 \Pi_l^2 }{ 1 - \xi^{2l+1} }\, \xi^{l+1}\,j_l(z_1)j_{l}(z_2)
\Biggr]
\label{FullereneDamping.19}
\end{eqnarray}
and
\begin{align}
\left\{
\begin{array}{l l}
\displaystyle{ S_{1l}(q) =  2 \Pi_l^4
{ \Bigl( \bigl(a_2-w_{1l}\bigr) j_{l}(z_2) + \xi^{l+1} w_{1l} j_l(z_1) \Bigr)^2
  \over (1-\xi^{2l+1})(a_2-w_{1l})(w_{2l} - w_{1l}) } }
\vspace{0.2cm} \\
\displaystyle{ S_{2l}(q) =  2 \Pi_l^4
{ \Bigl( \bigl(w_{2l}-a_2\bigr) j_{l}(z_2) - \xi^{l+1} w_{2l} j_l(z_1) \Bigr)^2
  \over (1-\xi^{2l+1})(w_{2l}-a_2)(w_{2l} - w_{1l}) }  } \ ,
\end{array} \right.
\label{FullereneDamping.20}
\end{align}
where $z_{1,2} = qR_{1,2}$, $w_{1l}$ and $w_{2l}$ are defined in Eq. (\ref{MultipoleVariation.2a}) and the following notations are used:
\begin{eqnarray}
\left\{
\begin{array}{l l}
\displaystyle{ a_1 = \frac{l}{\Pi_l^2} \bigl( 1-\xi^{2l+1} \bigr)   }
\vspace{0.2cm} \\
\displaystyle{ a_2 = \frac{l+1}{\Pi_l^2} \bigl( 1-\xi^{2l+1} \bigr) } \ .
\end{array} \right.
\label{FullereneDamping.21}
\end{eqnarray}

The functions $V_l(q)$, $S_{1l}(q)$ and $S_{2l}(q)$ are the diffraction factors depending on the transferred momentum $q$. They determine the relative significance of the multipole plasmon modes in various ranges of electron scattering angles.
Dominating contribution of different multipole modes results in the significant angular dependence of the differential electron energy loss spectrum \cite{Plasmons_experiment_2012}. This phenomenon arises due to the electron diffraction of the fullerene shell \cite{Mikoushkin_1998_PhysRevLett.81.2707}.
Considering a fullerene as an infinitely thin sphere, the dominating contribution of different multipole modes to the surface plasmon was shown in Refs. \cite{Gerchikov_1997_JPhysB.30.4133,Gerchikov_1998_JPhysB.31.3065}.


Since plasmons decay from the collective excitation mode to the incoherent sum of single-electron excitations, it is essential to account for the damping of plasmon oscillations. Therefore, one should introduce the finite widths, $\Gamma_l^{(v)}$ and $\Gamma_{jl}^{(s)}$ $(j = 1,2)$ of the volume and surface plasmon resonances, respectively,
and make the following substitutions in the right-hand side of Eq. (\ref{PRA.01}):
\begin{eqnarray}
\frac{1}{\om^2 - \om_{jl}^2} \longrightarrow \frac{1}{\om^2 - \om_{jl}^2 + \i\om\Gamma_{jl}^{(s)} } \ , \nonumber \\
\frac{1}{\om^2 - \om_p^2} \longrightarrow \frac{1}{\om^2 - \om_p^2 + \i\om\Gamma_l^{(v)} } \ .
\label{Damping.01}
\end{eqnarray}

Carrying out the imaginary parts produces:
\begin{eqnarray}
\Im \frac{1}{\om^2 - \om_{jl}^2 + \i\om\Gamma_{jl}^{(s)}} \longrightarrow
\frac{\om\Gamma_{jl}^{(s)} }{\bigl( \om^2 - \om_{jl}^2 \bigr)^2 + \om^2 \Gamma_{jl}^{(s)2} } \ , \nonumber \\
\Im \frac{1}{\om^2 - \om_p^2 + \i\om\Gamma_l^{(v)}} \longrightarrow
\frac{\om\Gamma_l^{(v)} }{\bigl( \om^2 - \om_p^2 \bigr)^2 + \om^2 \Gamma_l^{(v)2} } \ .
\end{eqnarray}
Thereby, the final formula for the differential inelastic scattering cross section with the account for three plasmons and with the damping included is the following:
\begin{equation}
\frac{\d^3\sigma}{\d\E_2 \d\Om_{{\bfp}_2}} = \frac{\d^3\sigma^{(v)} }{\d\E_2 \d\Om_{{\bfp}_2}} +
\frac{\d^3\sigma^{(s_1)} }{\d\E_2 \d\Om_{{\bfp}_2}} + \frac{\d^3\sigma^{(s_2)} }{\d\E_2 \d\Om_{{\bfp}_2}} \ ,
\label{FullereneDamping.17}
\end{equation}
where
\begin{eqnarray}
\left\{
\begin{array}{l l}
\displaystyle{
\frac{\d^3\sigma^{(v)} }{\d\E_2 \d\Om_{{\bfp}_2}} }
=
\frac{2 R_2 p_2}{\pi q^4 p_1} \, \om
\sum\limits_{l}
\frac{ \om_p^2\, \Gamma_l^{(v)} \, V_l(q) }{ \bigl(\om^2-\om_p^2\bigr)^2+\om^2\Gamma_l^{(v)2} }
\vspace{0.2cm}
\\
\displaystyle{
\frac{\d^3\sigma^{(s_1)} }{\d\E_2 \d\Om_{{\bfp}_2}} }
=
\frac{2 R_2 p_2}{\pi q^4 p_1} \, \om
\sum\limits_{l}
\frac{ \om_{1l}^2\,\Gamma_{1l}^{(s)} \, S_{1l}(q) }
{ \bigl(\om^2-\om_{1l}^2\bigr)^2+\om^2\Gamma_{1l}^{(s)2}}
\vspace{0.2cm} \\
\displaystyle{
\frac{\d^3\sigma^{(s_2)} }{\d\E_2 \d\Om_{{\bfp}_2}} }
=
\frac{2 R_2 p_2}{\pi q^4 p_1} \, \om
\sum\limits_{l}
\frac{ \om_{2l}^2\,\Gamma_{2l}^{(s)} \, S_{2l}(q) }
{ \bigl(\om^2-\om_{2l}^2\bigr)^2+\om^2\Gamma_{2l}^{(s)2} }
\end{array} \right.
\label{FullereneDamping.18}
\end{eqnarray}
and the functions $V_l(q)$, $S_{1l}(q)$ and $S_{2l}(q)$ are defined above.

Let us define the angular momentum range which is considered with the introduced model. In Ref. \cite{Gerchikov_1997_JPhysB.30.4133}, it was shown that excitations with large angular momenta $l$ have a single-particle nature rather than a collective character.
It follows from the fact that with increasing $l$ the wavelength of the plasmon mode 
becomes smaller than the characteristic wavelength of the delocalized electrons in the fullerene 
\cite{Gerchikov_1997_JPhysB.30.4133}.
In the case of the C$_{60}$ fullerene, only terms with $l \le 3$ should be included to the sum over $l$ in Eq. (\ref{FullereneDamping.18}), while multipole excitations with $l > 3$ are formed by single-electron transitions.

The introduced model is applicable within the long wavelength limit, when the characteristic scattering length, $1/q$, is large. Under the condition of the small transferred momentum $q$,
the volume plasmon is characterized by the constant frequency $\om_p$ which does not depend on the transferred momentum \cite{Landau_Lifshitz_10}.
In this paper, we do not consider the dependence of the plasmon widths on the transferred momentum which was studied in Ref. \cite{Gerchikov_2000_PhysRevA.62.043201}. The widths are treated as external parameters which are not calculated within the present model.

\section{Limiting cases of the general formulae
\label{LimitingCases}}

General expressions (\ref{FullereneDamping.17}) and (\ref{FullereneDamping.18}) for the differential inelastic scattering cross section are applicable for any spherically symmetric system with an arbitrary value of the ratio $\xi$.
In this section, we consider a transformation  of the general expressions in the two limiting cases: a metal cluster $(\xi = 0)$ and a fullerene modeled by an infinitely thin sphere $(\xi = 1)$.

\subsection{The 'metallic cluster limit'
\label{ClusterLimit}}

Metal cluster can be treated as a system with a uniform electron density distribution over the sphere of a radius $R$ \cite{Ekardt_1986_PhysRevB.29.1558}:
\begin{equation}
\rho_0(r) = \frac{N}{V}\, \Theta(R - r),
\end{equation}
where $V = 4\pi R^3/3$ is the cluster volume and $N$ is the number of delocalized electrons in a cluster.
In this limit, one obtains $\xi \to 0$, i.e.  $R_1\to 0$ and $R_2\equiv R$.

Electron density variation on the cluster surface leads to the formation of the surface plasmon, while the volume plasmon arises due to the density variation inside the system. In this case, the antisymmetric surface plasmon mode does not contribute to the cross section:
\begin{equation}
\lim_{\xi \to 0}\,
{\d^3 \sigma^{(s_2)} \over \d \E_2 \d \Om_{\bfp_2}} = 0
\label{ThinFullerene.11}
\end{equation}
and
the general expressions (\ref{FullereneDamping.17}) and (\ref{FullereneDamping.18}) for the differential cross section transform into the following:
\begin{equation}
\frac{\d^3 \sigma}{\d\E_2 \d\Om_{{\bfp}_2}} =
\frac{\d^3 \sigma^{(v)}}{\d\E_2 \d\Om_{{\bfp}_2}} + \frac{\d^3 \sigma^{(s)}}{\d\E_2 \d\Om_{{\bfp}_2}} \ ,
\label{ClusterLimit.9}
\end{equation}
where
\begin{eqnarray}
\frac{\d^3 \sigma^{(s)}}{\d\E_2 \d\Om_{{\bfp}_2}} &=& \frac {4 R\, p_2}{\pi q^4 p_1} \, \om
\sum_{l} \frac{ \Pi_l^4 \om_l^2\Gamma_l^{(s)} j_{l}^2(qR) }{ \bigl(\om^2-\om_l^2\bigr)^2+\om^2\Gamma_l^{(s)2} } \ ,
\label{ClusterLimit.10a}
\end{eqnarray}
\begin{eqnarray}
\frac{\d^3 \sigma^{(v)}}{\d\E_2 \d\Om_{{\bfp}_2}} &=& \frac {2 R^3 p_2}{\pi q^2 p_1} \, \om
\sum\limits_{l}
\frac{ \Pi_l^2 \om_p^2\Gamma_l^{(v)} F_l(qR) }{ \bigl(\om^2-\om_p^2\bigr)^2+\om^2\Gamma_l^{(v)2} }
\label{ClusterLimit.10b}
\end{eqnarray}
and
\begin{eqnarray}
F(z) = j_l^2(z) - j_{l-1}(z)j_{l+1}(z) - \frac2z j_{l}(z)j_{l+1}(z) \ .
\end{eqnarray}
Within this limit, the general expressions for the frequencies of the volume (\ref{Notation.4}) and the surface (\ref{MultipoleVariation.2a}) plasmons transform into the following: $\om_p = \sqrt{3 N/ R^3}$, \\ $\om_l = \sqrt{l/(2l+1)}\,\om_p$.

The differential cross section (\ref{ClusterLimit.9}) of the inelastic scattering on a metal cluster defined by Eqs. (\ref{ClusterLimit.10a}) and (\ref{ClusterLimit.10b}) coincides with expressions
presented earlier in Ref. \cite{Gerchikov_2000_PhysRevA.62.043201}.

\subsection{The 'infinitely thin fullerene' limit
\label{ThinFullerene}}

The model of a fullerene as an infinitely thin sphere was applied in a number of papers studying the photoionization \cite{Ivanov_2001_JPhysB.34.L669} and the electron scattering \cite{Gerchikov_1997_JPhysB.30.4133,Gerchikov_1998_JPhysB.31.3065} processes.
In this limit, one obtains $R_1\to R_2 \equiv R$ and $\xi \to 1$.

In the case of a sphere, the volume plasmon and the antisymmetric surface plasmon mode do not contribute to the cross section:
\begin{eqnarray}
\limxi\,
{\d^3 \sigma^{(v)} \over \d \E_2 \d \Om_{\bfp_2}}
=
\limxi\,
{\d^3 \sigma^{(s_2)} \over \d \E_2 \d \Om_{\bfp_2}}
=
0 \ .
\label{ThinFullerene.4c}
\end{eqnarray}
Therefore, the general expressions (\ref{FullereneDamping.17}) and (\ref{FullereneDamping.18}) reduces to:
\begin{equation}
{\d^3 \sigma \over \d \E_2 \d \Om_{\bfp_2}} =
\frac{4 R p_2}{\pi q^4 p_1} \, \om
\sum_{l}
{\Pi_l^4 \om_{l}^2\,\Gamma_{l}\, j_l^2(qR)
\over\bigl(\om^2-\om_{l}^2\bigr)^2+\om^2\Gamma_{l}^{2}}\, \ ,
\label{ThinFullerene.7}
\end{equation}
where
\begin{equation}
\om_{l} = \sqrt{\frac{l(l+1) N}{(2l+1) R^3}}
\end{equation}
  is the surface plasmon
frequency, and  $\Gamma_{l}\equiv \Gamma_{1l}^{(s)}$ is its
width.
Eq. (\ref{ThinFullerene.7}) coincides with the
expression presented earlier in Refs. \cite{Gerchikov_1997_JPhysB.30.4133,Gerchikov_1998_JPhysB.31.3065}.

\section{Conclusion
\label{Conclusion}}

In this paper, we have presented a detailed formalism for plasmon excitations in fullerenes caused by the collision with fast electrons. We have demonstrated that the energy loss spectrum is formed by three contributions, namely two modes of the surface plasmon and the volume plasmon.

Within the presented model, a fullerene was considered as a spherical shell of a finite width where the negative charge was distributed homogeneously over the shell. We showed that exposure of the system to an external field of a charged projectile causes the formation of the surface electron density variation on the surfaces of the shell and the volume density variation inside the shell. These variations lead to the formation of three plasmon excitations.

The introduced model can be applied to a spherically symmetric system with an arbitrary width of the shell. We have considered two limiting cases of the model when the system is treated as a charged full sphere ('metal cluster' limit) and as an infinitely thin sphere. The obtained expressions coincide with the results obtained earlier in Refs. \cite{Gerchikov_2000_PhysRevA.62.043201,Gerchikov_1997_JPhysB.30.4133,Gerchikov_1998_JPhysB.31.3065}.

Numerical results of the calculation based on the presented formalism as well as the comparison with the recent experimental data will be presented in another paper of this issue of the journal \cite{Plasmons_experiment_2012}. \\

A.V.V. is grateful to Deutscher Akademischer Austauschdienst (DAAD) for the financial support.

\appendix
\section{Interaction with a uniform external field
\label{Dipole_single-photon}}

In this Appendix, we consider the transformation of general expressions for the electron density variation obtained in Sec. \ref{Appendix_General_case} and \ref{Appendix_Finite_width} in the case of the uniform external field. This case describes the interaction with an electromagnetic field.

We assume that the wave length of electromagnetic radiation is much larger than the typical size of the system, i.e. the condition $\om R \ll 1$ is fulfilled. This condition implies the validity of the dipole approximation.

In this limit, $q = 0$ and Eq. (\ref{CS2002.Spherically.9c}) for the multipole variation of the electron density in a spherically symmetric system turns into the following one:
\begin{eqnarray}
\Bigl( \om^2 - 4\pi\rho_0(r) \Bigr) \delta\rho_l(r) &+& 4\pi\frac{\rho_0'(r)}{\Pi_l^2}
\int\limits_0^{\infty} g_l(r,r') \delta\rho_l(r') \d r'  \nonumber \\
& =& - \rho_0'(r) \phi_l'(r) \ ,
\label{Appendix.18}
\end{eqnarray}

Carrying out the transformations described in Sec. \ref{Appendix_Finite_width} and taking into account that in the dipole approximation the field intensity $\phi_l'$ does not depend on the spatial coordinate
\begin{equation}
\phi_l'(R_1) = \phi_l'(R_2) \equiv \phi_l' \ ,
\end{equation}
one derives the following equation:
\begin{align}
& \bigl( w - 1 \bigr) \delta\varrho_l(r) \Theta(r - R_1)\Theta(R_2 - r)  \nonumber \\
& \quad + \biggl(
w \sigma_l^{(1)} +
 I_l^{(1)} -
        \frac{l+1}{\Pi_l^2} \sigma_l^{(1)} + \frac{l}{\Pi_l^2} \sigma_l^{(2)}\, \xi^{l-1} \biggr)
 \delta(r - R_1)
\nonumber \\
& \quad + \biggl(
w \sigma_l^{(2)} +
 I_l^{(2)} +
        \frac{l+1}{\Pi_l^2} \sigma_l^{(1)} \xi^{l+2} - \frac{l}{\Pi_l^2}\sigma_l^{(2)} \biggr)
 \delta(r - R_2)
\nonumber \\
& \quad = - \frac{\phi_l'}{4\pi} \delta(r - R_1) + \frac{\phi_l'}{4\pi} \delta(r - R_2) \ .
\label{Appendix.19}
\end{align}

Matching the terms of different types on the right- and the left-hand side of Eq. (\ref{Appendix.19}), one obtains the equations for the volume and the surface electron density variations.
The equation for the volume density variation reads:
\begin{equation}
\bigl( w - 1 \bigr) \delta\varrho_l(r) \Theta(r - R_1)\Theta(R_2 - r) = 0 \ .
\end{equation}
Therefore, there is no volume plasmon arising in the system due to interaction with the uniform external field.

The total electron density variation is described only by the surface contributions:
\begin{equation}
\delta \rho_l(r) = \sigma_l^{(1)} \delta(r - R_1) + \sigma_l^{(2)} \delta(r - R_2) \ ,
\label{Appendix.20}
\end{equation}
which can be obtained solving the following system of equations:
\begin{eqnarray}
\left\{
\begin{array}{l l}
\displaystyle{ \biggl( w - \frac{l+1}{\Pi_l^2} \biggr) \sigma_l^{(1)} +
\frac{l}{\Pi_l^2}\xi^{l-1}\sigma_l^{(2)} = - \frac{\phi_l'}{4\pi} }
\vspace{0.2cm} \\
\displaystyle{ \frac{l}{\Pi_l^2}\xi^{l+2}\sigma_l^{(1)} +
\biggl( w - \frac{l}{\Pi_l^2} \biggr) \sigma_l^{(2)} = \frac{\phi_l'}{4\pi} }
\end{array} \right. \ .
\label{Appendix.21}
\end{eqnarray}

The solutions of Eq. (\ref{Appendix.21}) are
\begin{eqnarray}
\left\{
\begin{array}{l l}
\displaystyle{ \sigma_l^{(1)} = - \frac{1}{\Delta}\frac{\phi_l'}{4\pi} \left( w - \frac{l}{\Pi_l^2} \bigl( 1 - \xi^{l-1} \bigr) \right) }
\vspace{0.3cm} \\
\displaystyle{ \sigma_l^{(2)} = \frac{1}{\Delta}\frac{\phi_l'}{4\pi}  \left( w - \frac{l+1}{\Pi_l^2} \bigl( 1 - \xi^{l+2} \bigr) \right) } \ ,
\end{array} \right.
\label{Appendix.22}
\end{eqnarray}
where $\Delta = (w - w_{1l})(w - w_{2l})$ is the determinant of the system (\ref{Appendix.21}).

Therefore, one obtains the following expression for the total variation of electron density due to the surface plasmons:
\begin{eqnarray}
\delta\rho_l(r) &=& \displaystyle{ \frac{1}{\Delta}\frac{\phi_l'}{4\pi}
\left[
\left( w - \frac{l+1}{\Pi_l^2} \bigl( 1 - \xi^{l+2} \bigr) \right) \delta(r - R_2)  \right. } \nonumber \\
&-& \left. \left( w - \frac{l}{\Pi_l^2} \bigl( 1 - \xi^{l-1} \bigr) \right) \delta(r - R_1) \right] \ .
\label{Appendix.23}
\end{eqnarray}

For the further discussion let us introduce the multipole moment, $Q_l$, induced by the external field $\phi_l$:
\begin{equation}
Q_l = \frac{\sqrt{4\pi}}{\Pi_l} \int\limits_0^{\infty} r^{l+2}\, \delta\rho_l(r) \, \d r \ .
\label{NormalModes.1}
\end{equation}
Using (\ref{Appendix.23}) in (\ref{NormalModes.1}), one derives
\begin{equation}
Q_l = \frac{\sqrt{4\pi}}{\Pi_l} R_2^{l+2} \Sigma_l(\om) \ ,
\label{NormalModes.1a}
\end{equation}
where the following function is introduced:
\begin{align}
&\Sigma_l(w)
=
{1\over R_2^{l+2}}
\int\limits_0^{\infty} r^{l+2}\, \delta\rho_l(r) \, \d r
\nonumber\\
& \quad \quad \ =
\sigma_l^{(2)}(w) + \xi^{l+2}\sigma_l^{(1)}(w)
\nonumber\\
&= \frac{1}{\Delta}\frac{\phi_l'}{4\pi}
\left(
w(1 - \xi^{l+2} )
- {l+1 \over \Pi_l^2}
+  \xi^{l+2}
- {l \over \Pi_l^2} \, \xi^{2l+1}
\right) \ ,
\label{Appendix3.8}
\end{align}
and $\sigma_l^{(j)}(w) \equiv \sigma_l^{(j)} $, $j = 1,2$.

Let us rewrite the function $\Sigma_l(w)$ in the following form:
\begin{equation}
\Sigma_l(w) = \frac{\phi_l'}{4\pi} \frac{(w - w_0)}{\Delta} \, (1 - \xi^{l+2} ) \ ,
\end{equation}
where $w_0$ is defined as
\begin{eqnarray}
w_0
=
{(l+1) -  \Pi_l^2\ \xi^{l+2} + l\, \xi^{2l+1} \over \Pi_l^2(1 - \xi^{l+2} )} .
\label{Appendix3.9}
\end{eqnarray}

Performing some transformations and accounting for the expression for the determinant $\Delta$, one obtains the following expression:
\begin{align}
&\Sigma_l(w)
=
\frac{\phi_l'}{4\pi}
\left(
{ w_{1l} - w_0 \over w-w_{1l}}
-
{ w_{2l} - w_0 \over  w-w_{2l}}
\right)
{1 - \xi^{l+2} \over w_{1l} - w_{2l}} \ .
\label{Appendix3.10}
\end{align}
Finally, taking into account Eqs. (\ref{Notation.4}) and (\ref{MultipoleVariation.2a}), the function $\Sigma_l(w)$ can be represented in the following form which
clearly demonstrates the presence of two separate modes
characterized by the resonant frequencies $\om_{1l}$ and $\om_{2l}$:
\begin{eqnarray}
\Sigma_l(\om)
=
\frac{\phi_l'}{4\pi}
\left(
{ S_{1l} \over \om^2-\om_{1l}^2}
+
{ S_{2l} \over  \om^2-\om_{2l}^2}
\right) \ ,
\label{Appendix3.11}
\end{eqnarray}
where
\begin{eqnarray}
\left\{
\begin{array}{l l}
\displaystyle{
S_{1l}
=
\frac{3 N}{R_2^3}
{1 - \xi^{l+2} \over 1-\xi^3}\,
{ w_0 - w_{1l}  \over w_{2l} - w_{1l}} }
\vspace{0.2cm} \\
\displaystyle{
S_{2l}
=
\frac{3 N}{R_2^3}
{1 - \xi^{l+2}  \over 1-\xi^3}\,
{ w_{2l} - w_0 \over w_{2l} - w_{1l}} }
\end{array} \right.
\label{Appendix3.11a}
\end{eqnarray}
and $N$ is the number of delocalized electrons in the system.

Note that for the general expression of function $\Sigma_l(\om)$ one should account for the damping of plasmon oscillations and introduce the finite widths, $\Gamma_{1l}$ and $\Gamma_{2l}$, of the plasmon resonances according to Eq. (\ref{Damping.01}).

Due to interaction with the external electromagnetic field only dipole excitations may arise in the system. Therefore, the case of particular interest is $l=1$.

In the equations presented below, let us omit the second subscript allocated for $l$ in the notations for plasmon frequencies, $\om_1$ and $\om_2$, and their widths, $\Gamma_1$ and $\Gamma_2$. Below we consider only the case $l = 1$.

The resonant frequencies of the dipole surface plasmons are
\begin{eqnarray}
\left\{
\begin{array}{l l}
\displaystyle{
\om_{1}^2 =
{N \over 2 R_2^3}\,{ 3-p \over 1-\xi^3}
=
\frac{N}{ R_2^3}
{4\over (3+p)} }
\vspace{0.2cm} \\
\displaystyle{
\om_{2}^2 =
{N \over 2 R_2^3}\,{ 3+p \over 1-\xi^3}
=
\frac{N}{ R_2^3}
{4 \over 3-p} } \ ,
\end{array} \right.
\label{NormalModes_L1.3}
\end{eqnarray}
where the notation $p = \sqrt{1 + 8\xi^3}$ is introduced.

Calculating the values $w_0$, $w_{1l}$ and $w_{2l}$ for $l = 1$ and accounting for the plasmon widths, one obtains the expression for the dipole term of the function $\Sigma_l(\om)$:
\begin{align}
\Sigma_1(\om)
=
{3 N \phi'_1 \over 8\pi p\, R_2^3}
\left(
{ p + 1 \over \om^2-\om_{1}^2+ \i\,\om\Gamma_1} \right.
+
\left. { p - 1 \over  \om^2-\om_{2}^2+ \i\,\om\Gamma_2}
\right) .
\label{Damping.3}
\end{align}


The cross section of photoionization by a single photon is given by the general expression:
\begin{equation}
\sigma_1(\om) = {4\pi \om \over E}\, {\rm Im}\, D(\om) \ ,
\label{Photoionization.1}
\end{equation}
where $E$ is the strength of the external electric field,
and the induced dipole moment $D(\om)$ equals to $Q_1$ (\ref{NormalModes.1a}):
\begin{equation}
D(\om)
\equiv
\sqrt{4\pi\over 3}\,R_2^{3}\,\Sigma_1(\om)
\label{Photoionization.2}
\end{equation}
Choosing the potential $\phi_1(r)$ of the linearly polarized
 electromagnetic wave (in the dipole approximation) in the form
\begin{equation}
\phi_1(r) = - \sqrt{4\pi \over 3}\, r\, E
\label{Photoionization3.10}
\end{equation}
and using Eq. (\ref{Damping.3}), one derives
\begin{eqnarray}
 D(\om)
&=&
- E
{N \over 2p}
\nonumber \\
&\times&
\left(
{ p + 1 \over \om^2-\om_{1}^2+ \i\,\om\Gamma_{1}}
+
{ p - 1 \over  \om^2-\om_{2}^2+ \i\,\om\Gamma_{2}}
\right) .
\label{Photoionization.3}
\end{eqnarray}
The imaginary part of $D(\om)$ reads:
\begin{align}
&{\rm Im}\, D(\om)
=
E\,\om\,
{N \over 2p} \nonumber \\
&\times \left(
{ (p + 1)\Gamma_1 \over \bigl(\om^2-\om_{1}^2\bigr)^2+ \om^2\Gamma_1^2}
+
{ (p - 1)\Gamma_2 \over \bigl(\om^2-\om_{2}^2\bigr)^2+ \om^2\Gamma_2^2}
\right) \ .
\label{Photoionization.4}
\end{align}
Therefore, one evaluates the final expression for the photoionization cross section:
\begin{eqnarray}
\sigma_1(\om) &=&
4\pi \om^2
\left(
{ N_1\,\Gamma_1 \over \bigl(\om^2-\om_{1}^2\bigr)^2+ \om^2\Gamma_1^2} \right. \nonumber \\
&+&
\left. { N_2\,\Gamma_2 \over \bigl(\om^2-\om_{2}^2\bigr)^2+ \om^2\Gamma_2^2}
\right) \ ,
\label{Photoionization.6}
\end{eqnarray}
where
\begin{eqnarray}
N_1 = N\,{p + 1\over 2p} \ ,
\qquad
N_2 = N\,{p - 1\over 2p} \ .
\label{Photoionization.7}
\end{eqnarray}
This expression clearly shows that the photoionization cross section is defined by the two surface plasmons. These plasmons were observed at the experiment \cite{Scully_2005_PhysRevLett.94.065503} in interpretation made in Ref. \cite{Korol_AS_2007_PhysRevLett_Comment}.
The volume plasmon can be formed only when the system is exposed to a non-uniform external field, e.g. in collisions with charged particles.

\bibliography{EELS_Formalism}

\begin{thebibliography}{30}

\bibitem{Connerade_GR}
J.P. Connerade, J.M. Esteva, R.C. Karnatak, \emph{Giant Resonances in Atoms,
  Molecules, and Solids} (Plenum Publishing Corporation, 1987)

\bibitem{Brechignac_1989_ChemPhysLett.164.433}
C.~Brechignac, P.~Cahuzac, F.~Carlier, J.~Leygnier, Chem. Phys. Lett.
  \textbf{164}, 433 (1989)

\bibitem{Brack_1993_RevModPhys.65.677}
M.~Brack, Rev. Mod. Phys. \textbf{65}, 677 (1993)

\bibitem{Hertel_1992_PhysRevLett.68.784}
I.V. Hertel, H.~Steger, J.~de~Vries, B.~Weisser, C.~Menzel, B.~Kamke, W.~Kamke,
  Phys. Rev. Lett. \textbf{68}, 784 (1992)

\bibitem{Keller_Coplan_1992_ChemPhysLett.193.89}
J.W. Keller, M.A. Coplan, Chem. Phys. Lett. \textbf{193}, 89 (1992)

\bibitem{deHeer_1993_RevModPhys.65.611}
W.A. de~Heer, Rev. Mod. Phys. \textbf{65}, 611 (1993)

\bibitem{Kreibig_Vollmer}
U.~Kreibig, M.~Vollmer, \emph{Optical Properties of Metal Clusters}
  (Springer-Verlag, 1995)

\bibitem{Haberland}
H.~Haberland, \emph{Clusters of Atoms and Molecules: Theory, Experiment, and
  Clusters of Atoms}, Springer Series in Chemical Physics (Springer-Verlag,
  1994)

\bibitem{Ekardt}
W.~Ekardt, \emph{Metal Clusters}, Wiley Series in Theoretical Chemistry (Wiley,
  1999)

\bibitem{Solovyov_NATO_2001}
A.V. Solov'yov, in \emph{Atomic clusters and nanoparticles}, edited by C.~Guet,
  P.~Hobza, F.~Spiegelman, F.~David (EDP Sciences, Springer-Verlag,
  Berlin-Heidelberg-New York, 2001), p. 403

\bibitem{Solovyov_review_2005_IntJModPhys.19.4143}
A.V. Solov'yov, Int. J. Mod. Phys. B \textbf{19}, 4143 (2005)

\bibitem{Gerchikov_2000_PhysRevA.62.043201}
L.G. Gerchikov, A.N. Ipatov, R.G. Polozkov, A.V. Solov'yov, Phys. Rev. A
  \textbf{62}, 043201 (2000)

\bibitem{Bertsch_1991_PhysRevLett.67.2690}
G.F. Bertsch, A.~Bulgac, D.~Tomanek, Y.~Wang, Phys. Rev. Lett. \textbf{67},
  2690 (1991)

\bibitem{Ivanov_2001_JPhysB.34.L669}
V.K. Ivanov, G.Y. Kashenock, R.G. Polozkov, A.V. Solov'yov, J. Phys. B
  \textbf{34}, L669 (2001)

\bibitem{Madjet_2008_JPhysB.41.105101}
M.E. Madjet, H.S. Chakraborty, J.M. Rost, S.T. Manson, J. Phys. B \textbf{41},
  105101 (2008)

\bibitem{Reinkoester_2004_JPhysB.37.2135}
A.~Reink\"oster, S.~Korica, G.~Pr\"umper, J.~Viefhaus, K.~Godehusen,
  O.~Schwarzkopf, M.~Mast, U.~Becker, J. Phys. B \textbf{37}, 3125 (2004)

\bibitem{Scully_2005_PhysRevLett.94.065503}
S.W.J. Scully~\emph{et al}., Phys. Rev. Lett. \textbf{94}, 065503 (2005)

\bibitem{Korol_AS_2007_PhysRevLett_Comment}
A.V. Korol, A.V. Solov'yov, Phys. Rev. Lett. \textbf{98}, 179601 (2007)

\bibitem{Gerchikov_1997_JPhysB.30.4133}
L.G. Gerchikov, A.V. Solov'yov, J.P. Connerade, W.~Greiner, J. Phys. B
  \textbf{30}, 4133 (1997)

\bibitem{Gerchikov_1998_JPhysB.31.3065}
L.G. Gerchikov, A.N. Ipatov, A.V. Solov'yov, W.~Greiner, J. Phys. B
  \textbf{31}, 3065 (1998)

\bibitem{Mikoushkin_1998_PhysRevLett.81.2707}
L.G. Gerchikov, P.V. Efimov, V.M. Mikoushkin, A.V. Solov'yov, Phys. Rev. Lett.
  \textbf{81}, 2707 (1998)

\bibitem{Connerade_AS_PhysRevA.66.013207}
J.P. Connerade, A.V. Solov'yov, Phys. Rev. A \textbf{66}, 013207 (2002)

\bibitem{Lambin_Lukas_1992_PhysRevB.46.1794}
P.~Lambin, A.A. Lucas, J.P. Vigneron, Phys. Rev. B \textbf{46}, 1794 (1992)

\bibitem{Oestling_1993_EurophysLett.21.539}
D.~\"Ostling, P.~Apell, A.~Rosen, Europhys. Lett. \textbf{21}, 539 (1993)

\bibitem{Lo_Korol_AS_2007_JPhysB.40.3973}
S.~Lo, A.V. Korol, A.V. Solov'yov, J. Phys. B \textbf{40}, 3973 (2007)

\bibitem{Varshalovich}
D.A. Varshalovich, A.N. Moskalev, V.K. Khersonskii, \emph{Quantum Theory of
  Angular Momentum} (World Scientific Publishing, Singapore, 1988)

\bibitem{Kubo_lin_response_1962}
R.~Kubo, J. Phys. Soc. Japan \textbf{17}, 975 (1962)

\bibitem{Plasmons_experiment_2012}
P.~Bolognesi, L.~Avaldi, A.~Ruocco, A.~Verkhovtsev, A.V. Korol, A.V. Solov'yov,
  submitted to the current topical issue  (2012)

\bibitem{Landau_Lifshitz_10}
E.M. Lifshitz, L.P. Pitaevskii, \emph{Physical Kinetics: Volume 10}, Course of
  Theoretical Physics (Butterworth-Heinemann, 1981)

\bibitem{Ekardt_1986_PhysRevB.29.1558}
W.~Ekardt, Phys. Rev. B \textbf{29}(4), 1558 (1986)

\end{thebibliography}

\end{document}